\documentclass[12pt]{article}
\usepackage{amssymb,amsmath}
\usepackage{rotating}
\usepackage{mathrsfs}
\DeclareMathAlphabet{\pazocal}{OMS}{zplm}{m}{n}

\usepackage{curves}
\usepackage{epic}
\usepackage{eepic}
\usepackage{epsfig}

\newcommand{\dd}{\mbox{\rm d}}

\newcommand{\gam}{\gamma}

\newcommand{\dg}{\dagger}

\newcommand{\tl}{\tilde}

\newcommand{\DD}{\mbox{\rm D}}

\newcommand{\nnn}{\noindent}

\newcommand{\p}{\partial}

\newcommand{\be}{\begin{equation}}
\newcommand{\bear}{\begin{eqnarray}}

\newcommand{\srg}{\sqrt{-g}}

\newcommand{\ear}{\end{eqnarray}}
\newcommand{\ee}{\end{equation}}

\newcommand{\vs}{\vspace}
\newcommand{\hs}{\hspace}

\begin{document}

\begin{center}

\
\vs{1cm}

\baselineskip .7cm

{\bf \Large On the quantisation of gravity by embedding spacetime in a
higher dimensional space}\renewcommand{\thefootnote}{\fnsymbol{footnote}}\setcounter{footnote}{1}
\footnote{{\bf Note:} {\it This old paper was published in} Classical and Quantum
Gravity {\bf 2} (1985) 869--889; DOI: 10.1088/0264-9381/2/6/012. {\it The body of the posted
version is identical to the published one, except for corrections of few misprints.}
}

\vs{2mm}
\baselineskip .5cm
Matej Pav\v si\v c

Jo\v zef Stefan Institute, Jamova 39, SI-1000, Ljubljana, Slovenia; 

email: matej.pavsic@ijs.si

\vs{3mm}

\renewcommand{\thefootnote}{\arabic{footnote}}\setcounter{footnote}{0}

\end{center}

               {\bf Abstract}. Certain difficulties of quantum gravity can be avoided if we embed the spacetime
               $V_4$ into a higher dimensional space $V_N$; then our spacetime is merely a 4-surface in $V_N$.
               What remains is conceptually not so difficult: just to quantise this 4-surface. Our formal
               procedure generalises our version of Stueckelberg's proper time method of worldline
               quantisation. We write the equations of $V_4$ in the covariant canonical form starting from
               a model Lagrangian which contains the classical Einstein gravity as a particular case. Then
               we perform quantisation in the Schr\"odinger picture by using the concepts of a phase
               functional and wave functional. As a result we obtain the uncertainty relations which
               imply that an observer is `aware' either of a particular spacetime surface and has no
               information about other spacetime surfaces (which represent alternative histories); or
               conversely, he loses information about a particular $V_4$ whilst he obtains some information
               about other spacetimes (and histories). Equivalently, one cannot measure to an arbitrary
               precision both the metric on $V_4$ and matter distribution on various alternative spacetime
               surfaces. We show how this special case in the `coordinate' representations can be
               generalised to an arbitrary vector in an abstract Hilbert space.

\section{Introduction}

All attempts at the quantisation of general relativity have met so far with considerable
difficulties (see Anderson 1964, Brill and Gowdy 1970, Ashtekar and Geach 1974).
The problems are of various types, from the conceptual to the technical ones. So in
spite of the tremendous amount of work done on the subject we still do not possess
a generally accepted theory of quantum gravity. Forced with this undesirable situation
I am becoming inclined to the view that we have missed some important points in the
development of our concepts about the relation between quantum theory and general
relativity.

    Let me develop here my own view of the starting points of quantum gravity.
Relativistic quantum mechanics in itself is a paradox, as it unites `fatalism' (everything
is written in spacetime once and for all) with quantal probabilism (de Beauregard
1979). Since in the spacetime of general (or special) relativity all worldlines are `frozen'
and strict determinism is valid, there is no room for an observer's free decisions. This
is still true even if we postulate that an observer's `now' (by this we mean a
three-dimensional hypersurface $\Sigma$ of `simultaneous' events) moves forward along a certain
time-like direction in the spacetime $V_4$; the motions on $\Sigma$ are determined in advance.
In the moment when we postulate that a conscious observer has a free will (whatever
is meant by this) which is not an illusion but real we already have an inconsistency.
Instead of free will it is enough to take into account the validity of the quantum
probability principle and we find again that quantum mechanics is incompatible with
the existence of an objective reality in spacetime (Pav\v si\v c 1981a, b). In the case of
special relativity this incompatibility was not so obvious and we were still able to
construct more or less satisfactory relativistic quantum field theories. On the other
hand, in the case of general relativity this incompatibility is so fatal that, in my opinion,
it practically prevents any consistent construction of a quantum theory of gravity in
four-dimensional spacetime.

According to us this problem could be resolved---at least conceptually---if we embed
the spacetime $V_4$ into a certain higher dimensional space $V_N$; then our spacetime is
merely a 4-surface in a higher space. What remains for us to do is conceptually not
so difficult: just to quantise this surface. We already have some experience as to how
to quantise two-dimensional surfaces in spacetime (which are strings from a
three-dimensional point of view) (Polyakov 1981a, b, Tataru-Mihai 1982, Henneaux 1983,
Gervais and Neveu 1982, Fradkin and Tsetlin 1982). These methods or any other
analogous ones we have to generalise for the case of a four-dimensional surface.

Let me explain briefly how in higher space we avoid the incompatibility of the
concept of spacetime with the concept of `free will' or `free decision'. We may postulate
that higher space is pseudo Euclidean (essentially flat) and has a certain structure of
events (which are `physical' in a broader sense); this higher dimensional world is
deterministic, all events in it are frozen (these are either points, lines or a continuous
distribution of a certain `matter' density (see the following sections). Next, we can
postulate that what a conscious observer observes is a succession of certain
three-dimensional surfaces (also called the simultaneity surface or surface of `now') (Pav\v si\v c
 1981a, b); let us call this succession the motion of simultaneity 3-surface or $\Sigma$ motion.
 Some parts of this $\Sigma$ motion are under the direct conscious control of a certain observer:
he can move his arms and legs and thus influence the course of events on $\Sigma$. Other
parts of $\Sigma$ are out of an observer's direct control: he cannot influence the motion of
rivers, planets, etc. On a sufficiently small scale, the motion of $\Sigma$, even if outside an
observer's conscious or direct control, is due to quantum fluctuations, and therefore
 outside the predictive power of our observer. In all three cases, nevertheless, the $\Sigma$
 motion in higher space can have an arbitrary direction (within the constraints imposed
 by the theory to be described later), and describes a four-dimensional surface $V_4$. To
 different possible sequences of an observer's decisions\footnote{Various (in fact, many)
 possible streams of consciousness are flowing in a certain higher dimensional
structure $\bf B$; to a particular stream of consciousness there corresponds a particular detailed motion of $\Sigma$
through the structure $\bf B$. $\Sigma$ motion defines within $\bf B$ a four-dimensional structure which we call the brain.
To a certain $\Sigma_{\bf B}$ within $\bf B$ there corresponds a certain outside or external $\Sigma_{\rm out}$,
the correspondence being due
to the coupling of the external and the internal world through the sensory organs. This is closely related
to Wigner's point of view that the collapse of the wavefunction occurs in an observer's consciousness. See
 also the next three footnotes.
}
(or to different sequences of
 quantum decisions about the outcome of experiments) there correspond different
 surfaces $V_4$ which intersect different sets of events in higher space, thus bringing about
 different possible histories. All these various histories actually coexist in the higher
 space, but only one of them is followed by a given `stream of consciousness'\footnote{This is true
 only for particular experiments in which positions of events are measured.
 In general (e.g.
when energy and momentum are measured) a surface $V_4$ is not determined at all, as we shall see later,
and, a particular stream of consciousness chooses instead among the eigenstates (or `universes' according to the
Everett-Wheeler interpretation) in an abstract Hilbert space.
}.

    If $V_4$ is a curved 4-surface, then there is present a certain gravitational field on $V_4$.
Since $\Sigma$ motion fluctuates quantum mechanically at a certain microscopic scale, so
does the corresponding gravitational field.

In the following sections we shall develop a model theory based on the general
assumptions outlined here. The approach adopted seems to be free of the conceptual
and of some technical difficulties occurring in various previous approaches to quantum
gravity. So the difficulty with the non-arbitrariness of Cauchy data, related to the fact
that there are more variables in the theory than there are physical degrees of freedom
(see Brill and Gowdy 1970), does not occur anymore. Moreover, this approach also
resolves the interpretational difficulties and paradoxes of the conventional quantum
theory\footnote{This is implicit in the fact that the quantum theory of a spacetime surface in an embedding space is just
one particular representation of the Everett (1957)-Wheeler (1957,1973) interpretation of quantum mechanics
(see also DeWitt 1967b) in which the above mentioned interpretational paradoxes do not exist.}
 (objective reality, EPR paradox, Schr\"odinger's cat, measurement problem, etc).
All these difficulties apparently reduce to a single `difficulty', namely the acceptance
of a higher dimensional space with a given `matter' distribution, in which there exists
a conscious observer whose three-dimensional simultaneity 3-surface moves forward
in any direction according to a certain quantum law of motion\footnote{In our approach
we actually combine Wignerian dualism (Wigner 1967) with Everett-Wheeler-DeWitt
plurimundialism. Further details will exceed the scope of this paper and will
be given elsewhere. See also Pav\v si\v c (1981a, b).}.

\section{The postulates of quantum gravity}

The theory that we are going to develop is based on the following postulates.

    (i) There exists an $N$-dimensional space $V_N$ parametrised by the coordinates $\eta^a$
($a = 1, 2, 3, . . . , N$). In this space there exist static `material' or physical events
described by the matter density $\omega (\eta)$.

    (ii) There exists a three-dimensional surface $\Sigma$ moving in the space $V_N$. For an
observer, associated with a particular $\Sigma$ motion, of all material events in $V_N$ only those 
that lie on successive $\Sigma$ are observable.

    (iii) A simultaneity surface $\Sigma$ moves according to quantum laws. In the classical
limit its motion describes a four-dimensional continuum $V_4$, parametrised by the
coordinates $x^\mu$ ($\mu = 0, 1,2,3$) with the metric $g_{\mu \nu}$ and the matter
density $\omega ( x ) ) \equiv \rho ( x )$
satisfying the Einstein equations.

    As a working hypothesis we shall assume that the dimension $N$ of the embedding
space $V_N$ is 10; namely, according to the general theorem (Fronsdal 1959, Eisenhart,1926)
every $n$-dimensional Riemannian space can be embedded locally in a $N$-dimensional
pseudo-Euclidean space $M_N$ with $N = n ( n + 1)/2$, so that for $n = 4$ it is
$N = 10$. As stated already by Fronsdal (1959), all spacetimes which have been tested
experimentally so far\footnote{ We can hardly consider the Penrose plane wave spacetime (Penrose 1965) as a physical one, just because
no spacelike hypersurface exists for the global specification of Cauchy data.},
like the Schwarzschild solution, the Friedmann cosmological
solution, etc, can be embedded in $M_6$. So it seems reasonable to assume that a
ten-dimensional embedding space will suffice.

We shall not consider the complications which result from Clarke's (1970) work
which deals with global embedding of a generic spacetime (which moreover is not
necessarily a solution of Einstein's equations). In our approach we do not worry about
an embedding of a given spacetime; in other words, we do not start from the intrinsic
geometry of a $V_4$ and then search for its embedding, but on the contrary, we start
from the embedding space $M_N$---with a given dimension (say 10)---in which there
exists a 4-surface $V_4$. The latter satisfies a certain variational principle with respect to
$M_N$. Moreover, we consider $M_N$ as a physical space and not merely as an auxiliary
space; all events (with the coordinates $\eta^a$) of $M_N$ are physical, though classically an
observer is directly aware only of those events which belong to a certain spacetime
4-surface $V_4$. A given 4-surface $V_4$ is chosen by initial conditions and by the equation
of motion for $\eta^a(x)$.

In the following we shall first write the classical equations of motion of the surface
$\Sigma$; in other words, we shall write the equation of a four-dimensional surface $V_4$,
embedded in $V_N$. In principle the metric and curvature of $V_N$ can be arbitrary, but
we shall take it as a flat one. The metric on $V_4$ so obtained will satisfy the Einstein
equations.

    Next we shall quantise the motion of the 3-surface $\Sigma$. One possibility would be to
extend the procedure of the string quantisation (Bohr and Nielsen 1983, Kato and
Ogawa 1983): instead of a one-dimensional string in four-dimensional spacetime we
now have a three-dimensional surface in an $N$-dimensional space $V_N$. Instead of using
this approach, we shall rather follow a different though probably equivalent procedure
by formulating the theory in the Schr\"odinger picture.

\section{Equation of a classical $\Sigma$ motion: equation of a spacetime surface embedded in a
higher dimensional space}

We shall derive the equations of a surface $V_4$ from the following action:
$$
    W = \int \mathscr{L} \, \dd^4 x
\eqno{(3.1)}
$$
with the Lagrangian density
$$
   {\mathscr L} = \sqrt{-g} \, \left ( \frac{R}{8 \pi} + L_{\rm m} \right )
\eqno{(3.2)}
$$
where $R$ is the curvature scalar of $V_4$ and $L_{\rm m}$ is the matter Lagrangian
(to be specified latter).

If the surface $V_4$ embedded in a pseudo-Euclidean space $M_N$ is described
by the parametric equation
$$
  \eta^a = \eta^a (x) ~~~~~~(a=1,2,3,...,N)
  \eqno{(3.3)}$$
then the metric tensor on $V_4$ is given by
$$
  g_{\mu \nu}  = \p_\mu \eta^a \p_\nu \eta_a ~~~~~(\mu, \nu = 0,1,2,3).
\eqno{(3.5)}$$
The Riemann tensor takes the form  
$$
  R_{\mu \alpha \nu \beta} = \DD_\mu \DD_\beta \eta^a \DD_\alpha \DD_\nu \eta_a -
  \DD_\alpha \DD_\beta \eta^a \DD_\mu \DD_\nu \eta_a
\eqno{(3.5)}$$
where $\DD_\mu$ means the covariant derivative, so that (Eisenhart 1926) 
$$\DD_\mu \DD_\nu \eta_a = b_{ab} \p_\mu \p_\nu \eta^b 
  ~~~~~~~~~~b_{ab} =\delta_{ab} - \p_\rho \eta_a \p^\rho \eta_N
\eqno{(3.6)}$$
and where $\delta_{ab}$ is the diagonal pseudo-Euclidean metric tensor of the flat space $V_N$.
The curvature scalar is then
$$
   R=\DD^\mu \DD^\nu \eta^a \, \DD_\mu \DD_\nu \eta_a -
     \DD^\mu \DD_\mu \eta^a \, \DD^\nu \DD_\nu \eta_a .
\eqno{(3.7})
$$

We see that the Lagrangian density ${\mathscr L}$ given by (3.2) depends on the
variables $\eta^a$,
its first derivatives $\p_\mu \eta^a$ and second derivatives 
$\p_\mu \p_\nu \eta^a$:
$$
    {\mathscr L} = {\mathscr L} (\eta^a, \p_\mu \eta^a,\p_\mu \p_\nu \eta^a ).
\eqno{(3.2a)}
$$    
The variation principle $\delta W=0$ gives the following equation for
$\eta^a (x)$:
$$
   E_a \equiv \frac{\p {\mathscr L}}{\p \eta^a}
    - \p_\mu \frac{\p {\mathscr L}}{\p \p_\mu \eta^a}
    +  \p_\mu \p_\nu \frac{\p {\mathscr L}}{\p \p_\mu \p_\nu \eta^a} = 0.
\eqno{(3.8)}
$$    
In deriving equation (3.8) it was assumed that the surface $V_4$ is bounded
by a certain
three-dimensional surface on which the variations $\delta \eta^a$ and 
$\delta \p_\mu \eta^a$  are fixed and set to zero.

For the Lagrangian density (3.2) equations (3.8) assume the explicit form
$$
  \p_\mu \left [ \sqrt{-g} \left ( \frac{G^{\mu \nu}}{8 \pi} + T^{\mu \nu}
  \right ) \p_\nu \eta_a \right ] = 0
\eqno{(3.9)}
$$
where $G^{\mu \nu} \equiv R^{\mu \nu} - \frac{1}{2} g^{\mu \nu} R$ and
$T^{\mu \nu} = (1/\sqrt{-g})(\p L_{\rm m}/\p g_{\mu \nu} -
\p_\alpha \p L_{\rm m}/\p \p_\alpha g_{\mu \nu})$.

In order to further explore equation (3.9), let us introduce new variables
${C^\mu}_a$ and replace (3.9) by the system of equations
$$
  \frac{1}{\sqrt{-g}} \p_\mu (\sqrt{-g}) {C^\mu}_a ) = 0
\eqno{(3.10a)}
$$
$$
  \left ( \frac{G^{\mu \nu}}{8 \pi} + T^{\mu \nu} \right ) \p_\nu \eta_a
  = {C^\mu}_a .
\eqno{(3.10b)}
$$

Multiplying (3.10b) by $\p^\alpha \eta_a$ (and sum over a) one obtains
$$
   G^{\mu \nu} = - 8 \pi G (T^{\mu \nu} - C^{\mu \nu})
\eqno{(3.11)}
$$
with
$$  C^{\mu \nu} \equiv  {C^\mu}_a \p^\nu \eta^a .
\eqno{(3.12)}
$$

One can easily prove the relations
$$
   {C^\nu}_{a;\nu} = C^{\mu \nu} \DD_\mu \DD_\nu \eta_a 
   + {C^{\mu \nu}}_{;\nu} \p_\mu \eta_a
$$
$$
  {C^{\mu \nu}}_{;\nu}={C^{\nu}}_{a;\nu} \p^\mu \eta^a
$$
From (3.10a) it then follows ${C^{\mu \nu}}_{;\nu} = 0$. Since
${G^{\mu \nu}}_{;\nu} =0$ identically, we have also ${T^{\mu \nu}}_{;\nu} =0$.
Equations (3.11) are the Einstein equations, apart from the term 
$C^{\mu \nu}$ which
can be included in the redefinition of the stress-energy tensor $T^{\mu \nu}$.

By the way, since $G^{\mu \nu}$ and $T^{\mu \nu}$ are symmetric,
it follows also that $C^{\mu \nu}$ is symmetric.

    We see that the Lagrangian (3.2) which gives the Einstein equations when considered
as a function of the metric $g_{\mu \nu}$ and the derivatives $\p_a g_{\mu \nu}$
gives essentially the same
Einstein equations (apart from $C^{\mu \nu}$) also if it is considered as a function of the
embedding coordinates $\eta^a$  and the derivatives $\p_\nu \eta^a$, $\p_\mu \p_\nu \eta^a$.

   Let us now further investigate the properties of equation (3.9). By using the well
known relation
$$ \frac{1}{\sqrt{-g}} \p_\nu (\srg A^{\mu \nu})+ \Gamma^\mu_{\rho \nu} A^{\rho \nu}
  = {A^{\mu \nu}}_{;\nu}
$$
for a generic tensor $A^{\mu \nu}$ and the expression $\Gamma^\mu_{\alpha \beta} =
\p^\mu \eta^a \, \p_\alpha \p_\beta \eta_a$ for the affinity (Eisenhart                        
1926), we can write (3.9) in the equivalent forms
$$ \left ( \frac{G^{\mu \nu}}{8 \pi} + T^{\mu \nu} \right ) \DD_\mu \DD_\nu \eta_a= 0
\eqno{(3.9a)}
$$
$$ \frac{G^{\mu \nu}}{8 \pi} \DD_\mu \DD_\nu \eta_a + \frac{1}{\srg} \p_\mu (\srg \, T^{\mu \nu}
   \p_\nu \eta_a ) = 0.
\eqno{(3.9b)}
$$

   Since our Lagrangian (3.2) is invariant with respect to reparametrisation of the
coordinates $x^\mu$ on the subspace $V_4$ we expect four constraints satisfied by the equation
of motion; indeed using the identification (3.8) we have as a consequence of (3.9a) 
and (3.6):
$$  E_a \p^\nu \eta^a = 0.
\eqno{(3.13)}
$$
This identity is analogous to the well known identity $u^\mu \dd u_\mu/ds = 0$ which holds for
a free particle's worldline.

    Let us assume for the moment that ${\mathscr L}_{\rm m} \equiv \srg\, L_{\rm m} =
\srg \,\omega (1 - g_{\mu \nu} u^\mu u^\nu)$, where
$\omega (\eta)$ is an arbitrary function of position in $M_N$ (see later)
and $u^\nu$ is a certain unit
4-vector field on $V_4$. Then from (3.8) it follows that 
$T^{\mu \nu} = \omega (\eta (x)) u^\mu u^\nu$ which can be
identified with the dust stress-energy tensor, since we will consider 
$\omega (\eta(x)) \equiv \rho (x)$ as
the mass density in $V_4$, and $u^\mu$ as the 4-velocity field. Then the second term in (3.9b)
assumes the form
$$ \frac{1}{\srg} \p_\mu (\rho u^\mu u^\nu \p_\nu \eta_a \srg ) =
  \frac{1}{\srg} \p_\mu ( \srg \, \rho u^\mu U_a ).
\eqno{(3.14a)}
$$
Since (3.9b) implies $\p^\mu \eta_c \,\p_\mu (\omega U^c) = 0$ which is identical to
 the conservation of rest
mass $(1/\srg) \p_\mu (\srg\, \rho u^\mu) = 0$, it can be written as
$$  
 \frac{G^{\mu \nu}}{8 \pi} \DD_\mu \DD_\nu \eta_a + \omega \,\dd U^a/\dd s = 0
\eqno{(3.14b)}
$$
where $U_a = \p^\nu \eta_a u_\nu$ is the velocity $U_a =\dd \eta_a/\dd s$ with
respect to $M_N$ and $\dd U_a/\dd s = u^\mu \p_\mu (u^\nu \p_\nu \eta_a)$. 
 By the way, $u^\mu = \dd x^\mu/\dd s = U^a \p^\mu \eta_a$ and
 $\dd s^2 = \dd x^\mu \dd x_\mu = \dd \eta^a \dd \eta_a$. Let 
us multiply (3.14b) by $\p^\alpha \eta^a$ and sum over $a$; one immediately observes
that $\DD_\mu \DD_\nu \eta_a \p^\alpha \eta^a = 0$ so that it remains
$$  
  u^\mu \p_\mu (u^\nu \p_\nu \eta_a) \p^\alpha \eta^a = 
  \frac{\dd U_a}{\dd s} \p^\alpha \eta^a = \frac{\dd u^\alpha}{\dd s}
  + \p^\alpha \eta^a \p_\mu \p_\nu \eta_a u^\mu u^\nu = 0
$$
which is the geodesic equation.

    We have seen that from our second-order Lagrangian ${\mathscr L} =
{\mathscr L}(\eta^a,\p_\mu \eta^a, \p_\mu \p_\nu \eta^a )$ we
obtain equation (3.9a)
which is not a fourth-order equation (as expected) but merely
a second-order equation. This is in agreement with the result obtained by Rund (1971)
who extensively studied variational problems on subspaces of a Riemannian manifold.
The occurrence of a second-order equation of motion from a second-order Lagrangian
indicates that our special Lagrangian entails some kind of degeneracy (Rund 1971)
which results in the non-uniqueness of the solution $\eta^a (x)$ to the preceding variational
procedure. In order to fix a solutionv $\eta^a (x)$ we need some additional equation. There
are certainly various possible ways of completing the equations (3.9) or (3.9a). Here
I shall tentatively adopt the following procedure.

First, let us observe that our Lagrangian (3.2) for dust,
${\mathscr L} = \srg ( R/8 \pi + \omega (1-g_{\mu \nu} u^\mu u^\nu))$,
in the case when the Einstein equations are satisfied,
becomes ${\mathscr L} = \srg \, \omega$. Let us denote the latter Lagrangian by
${\mathscr L}_1 = {\mathscr L}_1 (\eta^a, \p_\mu \eta^a)$ and
the former one by ${\mathscr L}_2 = {\mathscr L}_2 (\eta^a, \p_\mu \eta^a, \p_\mu \p_\nu \eta^a)$.
From the Lagrangians ${\mathscr L}_1$ and ${\mathscr L}_2$ we obtain
the following system of equations:
$$
  \frac{1}{\srg} \p_\mu (\srg\, \omega \p^\mu \eta_a) = \frac{\p \omega}{\p \eta^a}
\eqno{(3.14)}
$$
$$
 \frac{1}{\srg} \p_\mu (\srg\, \omega u^\mu U_a) 
 = - \frac{1}{8 \pi} G^{\mu \nu} \DD_\mu \DD_\nu \eta_a .
\eqno{(3.15)}
$$
The quantity $\omega (\eta)$ is the matter density in the higher space and is a given
function of
$\eta^a$. The choice of $\omega (\eta^a)$ depends on the model of the universe
that we adopt and the
scale which we are interested in. For given initial and boundary values of $\eta^a$ and for
a chosen parametrisation $x^\mu$ we can calculate $\eta^a (x)$ from (3.14).
Once $\eta^a (x)$ is known,
we know also $g_{\mu \nu}$ and $G^{\mu \nu}$. Therefore in (3.15) only the velocity
$U_a (x)$ of a point on
$V_4$ is unknown (remember that $u_\mu = \p_\mu \eta^a U_a$) and can be calculated
from the equation.
In other words, though $V_4$ is known from (3.14) and is represented by the equation
$\eta^a = \eta^a(x)$, we still do not know the direction
(i.e. $U^a$ or $u_\mu =\p_\mu \eta_a U^a$)  into which
the mass flows, unless we solve (3.15). We have also seen that (3.15) automatically
implies the conservation of rest mass $\DD_\mu (\rho u^\mu)=0$ and the validity of the Einstein
equations (3.11) (with addition of the term $C^{\mu \nu}$) together with 
$\DD_\nu (\rho u^\mu u^\nu)=0$ and $\DD_\nu C^{\mu \nu} = 0$.

However, it may happen that $C^{\mu \nu} = 0$. In another paper (Pav\v si\v c: 1985) we have an
example of $\omega (\eta)$, $\eta^a (x)$ and $U^a$  which solve our system (3.14) and (3.15).
     
From the preceding we can conclude that our starting Lagrangian for the `field'
$\eta^a (x)$ can be taken to be ${\mathscr L} = \srg\, \omega (\eta)$
 with the corresponding equation of motion $\DD_\mu (\omega \p^\mu \eta_a) 
 = \p \omega/\p \eta_a$  (equation (3.14)). The 'true' dynamical variables in our theory are
the coordinates $\eta^a (x)$ of the spacetime surface $V_4$. The metric tensor components
$g_{\mu \nu}$ are not `true' dynamical variables. In the usual approaches to quantum gravity
they caused troubles, since $g^{\mu \nu}$ are not all independent but are related through
${G^{\mu \nu}}_{;\nu}= 0$.
Moreover, one cannot arbitrarily specify Cauchy data, namely $g_{ij} (x)$ and
$\pi^{k l} (x)$, since
they must satisfy the Einstein equations $G^{0 \mu}= 0$. On the other hand, $\eta^a (x)$ and
$\pi^a (x)$  can be specified at will on a given 3-surface $\Sigma$. Though the variables 
$\eta^a (x)$ and $\pi^a (x)$ are not
all independent, but obey the four identities (3.13), this does not cause any trouble in
setting the theory into a canonical form. Thus, this is analogous to the situation which
occurs in special relativity, where a worldline is described by $x^\mu = x^\mu (\lambda)$,
$\lambda$ being an
arbitrary parameter. In the following we shall make the canonical formulation of the
theory of a spacetime sheet satisfying this specific first-order Lagrangian.

\section{Generator for infinitesimal transformations, momentum and stress-energy tensor of
the field}

We may consider $\eta_a (x)$ as a field entering the Lagrangian density ${\mathscr L}$.
Let us confine us to a first-order Lagrangian, say
$$ {\mathscr L} = \srg\, \omega (\eta)
\eqno{(4.1)}
$$
and let $W= \int {\mathscr L} \, \dd^4 x$ be the action. The variation of $W$ is
$$  \delta W = \int \left ( \frac{\p {\mathscr L}}{\p \eta^a} \delta \eta^a +
  \frac{\p {\mathscr L}}{\p \p_\mu \eta^a} \delta \p_\mu \eta^a \right ) \dd^4 x .
\eqno{(4.2)}
$$
This can be rearranged so that after taking into account the equations of motion
$\p {\mathscr L}/\p \eta^a - \p_\mu \p {\mathscr L}/\p \p_\mu \eta^a = 0$ we obtain
$$  \delta W = \int \p_\mu \left ( \frac{\p {\mathscr L}}{\p \p_\mu \eta^a} \delta \eta^a
  \right ) \dd^4 x .
\eqno{(4.3)}
$$
This represents the variation of the action $W$ when going from a spacetime surface
$V_4$ to another spacetime surface $V_4+ \delta V_4$ (see figure 1a), where both $V_4$
and $V_4+ \delta V_4$
are solutions of the field equations, explicitly of equation (3.14).

\setlength{\unitlength}{.8mm}

\begin{figure}[h!]
\hs{3mm}
\begin{picture}(120,70)(-10,-95)
\begin{turn}{-90}
\put(0,0){\includegraphics[scale=0.60]{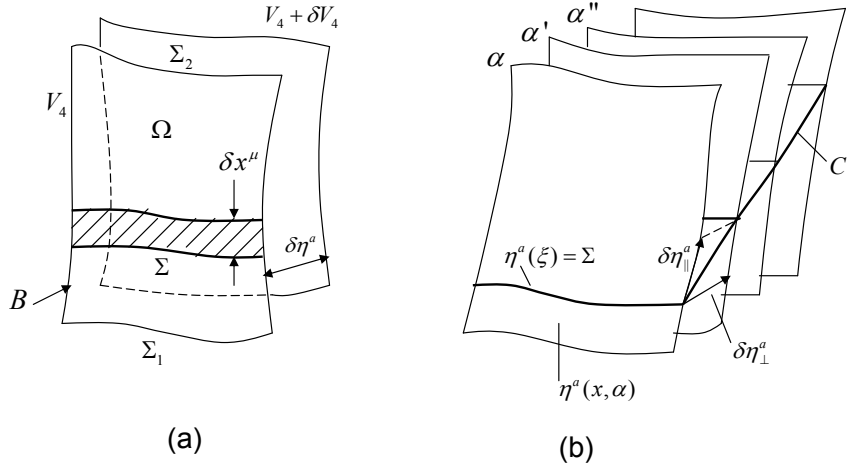}}
\end{turn}
\end{picture}

\caption{\footnotesize (a) The variation of the action W between two successive spacetime slices $V_4$
and $V_4+ \delta V_4$. (b) A family of spacetime slices and the path $C$ of $\Sigma$.}
\end{figure}

Let us introduce the notation
$$  {\pi^\mu}_a = \frac{\p {\mathscr L}}{\p \p_\mu \eta^a}
\eqno{(4.4)}
$$
and let us express $\delta W$ in (4.3) as the integral over a closed three-dimensional surface
${\bf B}$ confining the segment $\Omega$ of $V_4$:
$$
  \delta W = \oint \dd \Sigma_\mu \, {\pi^\mu}_a \delta \eta^a
\eqno{(4.5)}
$$
where $\dd \Sigma_\mu$ is a 3-surface element. Now we assume that the derivatives 
$\p_\mu \eta^a$ and $\p {\mathscr L}/\p \p_\mu \eta^a$
tend to zero at spatial infinity (Barut and Mullen 1962,\,1964)\footnote{
Alternatively, we may assume that spacetime $V_4$ is spatially closed.}.
We can then
assume that ${\bf B}$ consists of two spacelike 3-surfaces $\Sigma_1$ and $\Sigma_2$
alone (figure la). Equation (4.5) becomes
$$
  \delta W = \int_{\Sigma_1}^{\Sigma_2} \dd \Sigma_\mu \, {\pi^\mu}_a
  \delta \eta^a .
\eqno(4.6)
$$
The expression
$$
  \delta S_1 = \int \dd \Sigma_\mu \, {\pi^\mu}_a \delta \eta^a = G_1 (\Sigma)
\eqno(4.7)
$$
we shall call the generator of variations of $V_4$ satisfying the equations of motion.

Let $\xi_r ~ (r = 1,2,3)$ be the coordinates which parametrise the 3-surface $\Sigma$.
By writing
the 3-surface element $\dd \Sigma$ as the product of the three-dimensional (spatial) volume
$\dd^3 \xi$ and the normal $n^\mu$
$$
   \dd \Sigma_\mu =\dd^3 \xi n_\mu = \dd \Sigma \, n_\mu
\eqno{(4.8)}
$$
and by identifying
$$
   \pi_a \equiv {\pi^\mu}_a n_\mu
\eqno{(4.9)}$$
equation (4.7) becomes
$$
   \delta S_1 = \int_\Sigma \dd \Sigma\, \pi_a \delta \eta^a .
\eqno{(4.10)}
$$
In the case of our Lagrangian (4.1) we obtain from (4.4) and (4.9)
$$
   {\pi^\mu}_a = \srg\, \omega \,\p^\mu \eta_a
\eqno{(4.11)}$$
$$
   \pi_a = \srg\, \omega \,\p^\mu \eta_a n_\mu .
\eqno{(4.12)}$$
Further, we can choose $\Sigma$ such that the normal $n_\mu$ is equal to the 4-velocity
$n_\mu = u_\mu$.
Then we obtain
$$
  \pi_a =\srg\, \omega U_a
\eqno{(4.13)}
$$
where $U_a = \p_\mu \eta_a u^\mu$ is the velocity of a point on $V_4$ with
respect to the embedding space.

So far we have considered the variation of the action $W$ such that the boundary
of integration on a given $V_4$ was fixed but the variations on the boundary were different
from zero. Now we extend the variation of $W$ so that the boundary is also subjected
to variation. This induces the variation of the coordinates $x^\mu$ of the boundary into
the coordinates $x'^\mu = x^\mu + \delta x^\mu$ of a new boundary. The contribution of this variation is
$$
  \delta_B \int {\mathscr L}\, \dd^4 x = \int_{\Omega-\Omega'} {\mathscr L} \,\dd^4 x =
  \oint_{\bf B} {\mathscr L}\, \dd \Sigma_\mu \, \delta x^\mu =
  \int_{\Sigma_1}^{\Sigma_2} {\mathscr L}\, \dd \Sigma_\mu \, \delta x^\mu
\eqno{(4.14)}
$$
where in the last step we assumed that the boundary surface ${\bf B}$ consists of the initial
and the final surfaces $\Sigma_1$ and $\Sigma_2$ alone. The generator of this variation is then
$$
   \delta S_2 = \int_\Sigma \dd \Sigma_\mu \, {\mathscr L} \, \delta x^\mu 
   = G_2 (\Sigma) .
\eqno{(4.15)}
$$

The total generator is then the sum of (4.6a) and (4.15):
$$
   \delta S = \delta S_1 + \delta S_2 \equiv G(\Sigma) .
\eqno{(4.16)}
$$
This last expression can be written as
$$
  \delta S = \int \dd \Sigma (\pi_a\, \delta \eta^a (x) + {\mathscr L}\, \delta s )
\eqno{(4.17)}
$$
where $\dd \Sigma_\mu = \dd^3 \xi \,n_\mu$, $\delta x^\mu = n^\mu\delta s (\xi)$, $n^\mu$ being
the normal vector to $\Sigma$, and $\xi \equiv \xi^r$ 
the 3-coordinates on $\Sigma$. Occasionally we shall call $s(\xi) = \int \delta s (\xi)$
 the proper time
function; this name is justified by the fact that if we choose $\Sigma$ such that the normal
$n^\mu$ coincides with the 4-vector $u^\mu$, then the vector $\delta x^\mu (\xi) =
u^\mu \delta s (\xi)$ is tangential to a
geodesic characterised by the parameters $\xi^r~~ ( r = 1,2,3)$. Here $\delta \eta^a$
is the following variation of $V_4$:
$$
   \delta \eta^a = \eta'^a (x) - \eta^a (x) .
\eqno{(4.18)}
$$
If $\eta^a (x)$ is considered as a field, then this variation is identical
to the intrinsic variation of the field.

Now we can consider the total variation
$$
  {\bar \delta} \eta^a (x) = \eta'^a (x') - \eta^a (x) = \delta \eta^a (x) +
  \p_\nu \eta^a \delta x^\nu .
\eqno{(4.19)}
$$
Then the generator can be written as
$$
  \delta S = \int \dd \Sigma_\mu ({\pi^\mu}_a {\bar \delta} \eta^a (x) - {{\mathscr T}^\mu}_\nu 
  \delta x^\nu )
\eqno{(4.20)}
$$
where
$$
  {{\mathscr T}^\mu}_\nu = {\pi^\mu}_a \p_\nu \eta^a
   - {\mathscr L} {\delta^\mu}_\mu
\eqno{(4.21})
$$
is the stress-energy tensor. This formal stress-energy tensor should not be confused
with the one in the Einstein equations. If the Lagrangian is given by (3.2) or (4.1)
then the stress-energy tensor is explicitly
$$
  {\mathscr T}_{\mu \nu} = \srg\, (\omega \p_\mu \eta_a \,\p_\nu \eta^a -
  \omega g_{\mu \nu})
\eqno{(4.22)}
$$
which is equal to zero. Therefore (4.20), as a consequence of 
${\bar \delta} \eta^a (x)|_\Sigma = \delta \eta^a (\xi)$ (see
appendix 1) is simply
$$
  \delta S = \int \dd \Sigma_\mu \,{\pi^\mu}_a {\bar \delta} \eta^a (x) =
  \int \dd \Sigma \,\pi_a {\bar \delta} \eta^a (x) =
  \int \dd \Sigma \, \pi_a \delta \eta^a (\xi) .
\eqno{(4.23})
$$
This can be considered as a variation of the so-called phase functional
$S[\eta (\xi)]$ defined
in appendix 1. The functional derivative of $S$ with respect to
$\eta^a (\xi)$ is  the canonical momentum
$$
  \pi_a (\xi) = \frac{\delta S}{\delta \eta^a (\xi)}= \pi_a (\xi)[\eta(\xi)] .
$$

From equation (4.21) we obtain the following Hamiltonian functional:
$$
  H = \int \dd \Sigma_\mu \, n_\nu {\mathscr T}^{\mu \nu}
   = \int \dd \Sigma(\pi_a U^a - {\mathscr L}) \equiv \int \dd \Sigma \, {\mathscr H} ~~~~~~~~
   (U^a = n^\nu \p_\nu \eta^a)
\eqno{(4.24)}
$$
where $ n^\mu \delta s = \delta x^\mu$.

    Though the Hamiltonian is zero we can still use its functional dependence on
$\pi_a$ and $U^a$ to derive the Hamiltonian equations of motion.
Namely, by separating the
derivative a into a normal (to the 3-surface element $\dd \Sigma_\mu$)
directional derivative ${\hat \p}$ and a tangential derivative ${\tl \p}_\mu$
as follows (Barut and Mullen 1962, 1964)
$$
  \p_\mu = n_\mu {\hat \p} + {\tl \p}_\mu~~~~~~~~~~{\hat \p} \equiv n_\mu \p^\mu
\eqno{(4.25)}
$$
one can show that a variation of the Hamiltonian is
$$
  \delta H = \int \dd \Sigma \left ( {\hat \p} \eta^a \delta \pi_a -
  {\hat \p} \pi_a \delta \eta^a (\xi) \right ).
\eqno{(4.26)}
$$
From (4.26) we obtain the following equations of motion in the Hamiltonian form\footnote{
It is the 3-surface $\Sigma$ which moves.}
for the canonical variables $\pi_a (\xi)$ and $\eta^a (\xi)$:
$$
  {\hat \p} \pi_a = - \frac{\delta H}{\delta \eta^a (\xi)}=  \lbrace \pi_a, H \rbrace
  ~~~~~~~~~~~~~~{\hat \p} \eta^a = \frac{\delta H}{\delta \pi_a} 
  = \lbrace \eta^a,H \rbrace
\eqno{(4.27)}
$$
where
$$
  \lbrace u , v \rbrace = \frac{\delta u}{\delta \eta^c} \frac{\delta v}{\delta \pi_c}
  - \frac{\delta u}{\delta \pi_c} \frac{\delta v}{\delta \eta^c}
$$
is the Poisson bracket. These equations (4.27) are equivalent to the field equation (3.14).

Incidentally, we observe that the generator (4.20) can be written as
$$
  \delta S = \int \dd \Sigma (\pi_a {\bar \delta} \eta^a (x) - {\mathscr H}
  \delta s) = \int \dd \Sigma (\pi_a {\delta} \eta^a (\xi) - {\mathscr H}
  \delta s)
$$
The partial functional derivative of $S$ with respect to $s$ is
$$
  \frac{\delta S}{\delta s} = \frac{\delta_{\rm P} S}{\delta s} =
  - {\mathscr H} = 0 .
\eqno{(4.28)}
$$
This derivative I call partial, because the functional dependence on $s$ is also included
in the total variation ${\bar \delta} \eta^a (x)$. The total and the partial functional derivative are related
according to
$$
   \frac{\delta_{\rm T} S}{\delta s} = \frac{\delta_{\rm P} S}{\delta s} +
   \frac{\delta S}{\delta \eta^a (\xi)} \frac{\dd \eta^a}{\dd s}=
  - {\mathscr H} + \pi_a {\hat \p} \eta^a = {\mathscr L}
\eqno{(4.29)}
$$
which is consistent with the definition of ${\mathscr H}$ given by (4.24).

Having defined the generator for infinitesimal transformations, momentum and the
stress-energy tensor of the field $\eta^a (x)$ (which is actually a spacetime surface $V_4$), we
are already prepared to perform the formal quantisation of the theory.

\section{Quantisation of a spacetime surface embedded in a higher dimensional space}

We could quantise the motion of a three-dimensional surface $\Sigma$ in a higher dimensional
(more than four) space by using techniques analogous to those used in the
quantisation of strings (Polyakov 1981a, b, Fradkin and Tsetlin 1982, Horwitz and
Piron 1973, Horwitz and Arshanski 1982, Menski 1976, Aghassi et al 1970). String is
a one-dimensional continuum moving in a (e.g.) four-dimensional space thus describing
a two-dimensional continuum $V_2$. Here we wish to demonstrate the basis of another
method which appears suitable to a direct understanding of quantum gravity. We shall
generalise our version (Pav\v si\v c 1984) of the Stueckelberg (1941a, b, c) proper time
method of worldline quantisation, i.e. the quantisation of the motion of a zero-
dimensional 'continuum' ---a point which describes a one-dimensional continuum, a
worldline. Instead of a one-dimensional classical continuum, a worldline, we now
have a four-dimensional classical continuum, a spacetime surface.

We shall assume that a quantum state corresponding to a surface $V_4$ can be
represented by a wave functional $\psi(\Sigma,s)\equiv 
\psi[\eta(\xi),s(\xi)]$ where $\eta^a (\xi)$ is the
parametric equation (in $V_N$) of the simultaneity surface $\Sigma$,
and $s = s(\xi)$ is the proper time function defined in \S \,3.

In order to find the equation for a wave functional let us proceed as follows. First,
let us replace the Hamiltonian (4.24) by the Hamiltonian operator
$$
  {\hat H} = \int \dd \Sigma ({\hat \pi}_a {\hat U}^a - {\hat {\mathscr L}}) 
  \equiv \int {\hat {\mathscr H}} \dd \Sigma .
\eqno{(5.1)}
$$
In a suitable representation we can set
$$
  {\hat \pi}_a = - i \frac{\delta}{\delta \eta^a (\xi)}~~~~~~~~~
  {\hat U}^a = \gam^a ~~~~~~~~~~~{\hat {\mathscr L}} = -i \frac{\delta_{\rm T}}{\delta s}
   ~~~~~~~~{\hat {\mathscr H}} = i \frac{\delta}{\delta s}
\eqno{(5.2)}
$$
where it will turn out (see equation (5.7)) that the velocity operator ${\hat U}^a$  can be
represented by the Dirac matrices satisfying ($\delta_{ab}$ is a pseudo-Euclidean metric of
$V_N = M_N$):
$$
  \gam_a \gam_b + \gam_b \gam_a = 2 \delta_{ab} .
\eqno{(5.3)}
$$
Then we observe that classically $\delta S/\delta s = H=0$, and we set the analogous quantum
equation
$$
  i \frac{\delta \psi}{\delta s} = {\hat {\mathscr H}} \psi [\eta(\xi),s(\xi)] =
  ({\hat \pi}_a \gam^a - {\hat{\mathscr L}}) \psi = 0 .
\eqno{(5.4)}
$$
This equation is a generalisation of Dirac's equation. We also observe that (5.4) is a
kind of Tomonoga-Schwinger equation (see Blokintsev 1973) which in turn is a
generalisation of the Schr\"odinger equation. It is now understood that $\psi[\eta(\xi),s(\xi)]$
is a functional multiplied by a suitable spinor.

Equation (5.4) implies that there is no evolution along the proper time
function $s(\xi)$:
$$
  \psi[\eta(\xi),s(\xi)] = {\rm exp} \left ( - i \int {\hat H} \delta s \right )
  \psi[\eta(\xi),0] = \psi[\eta(\xi),0]
\eqno{(5.5)}
$$
so that matrix elements or expectation values of operators remain constant with $s(\xi)$
(see also DeWitt 1967a, b). However they do change with $\eta^a (\xi)$; and since the totality
of $\eta^a (\xi)$ fill all the space $V_N$, it still holds that matrix elements change with `time'.
Time is now not the proper time $s(\xi)$  but $\Sigma$ (i.e. $\eta^a (\xi)$)
(see also Pav\v si\v c (1984) and \S 6).

A particular solution---with definite $\pi_a$ and ${\mathscr L}$---of the wave functional equation
(5.4) is
$$
  \psi_\pi [\eta(\xi)] 
  = \psi_0\, {\rm exp} \left ( i \iint \pi_a \delta \eta^a (\xi)\, \dd \Sigma \right )
\eqno{(5.6)}
$$
under the condition
$$
  (\pi_a \gam^a )^2 = \pi_a \pi^a = {\mathscr L}^2
\eqno{(5.7)}
$$
where ${\hat {\mathscr L}} \psi_\pi= {\mathscr L} \psi_\pi$. In equations (5.6) and (5.7) 
$\pi^a (\xi)=\pi^a (\xi)[\eta (\xi)]$ and ${\mathscr L} (\xi) = {\mathscr L} (\xi) [\eta(\xi)]$
are the eigenvalue fields\footnote{
In a general case the eigenvalues are not constants but fields (see Pav\v si\v c 1982).}
corresponding to the operators (5.2). These
(functional) fields are defined over the set $\lbrace \eta^a(\xi): \eta^a(\xi) \subset V_4$,
$V_4 \in \ \lbrace V_4 \rbrace \rbrace$,
i.e. the set of $\eta^a (\xi)$ that  belong to the family of $V_4$
defining the particular $\pi_a$.
             
Each $\eta^a(\xi)$ represents a certain geometry. Wheeler (1967) represented a given
3-geometry by a point in the so-called superspace, let us call it $g$ superspace.
Analogously, we can consider a given 3-surface as a point in the corresponding $\Sigma$
superspace which is analogous to the g superspace, but not identical.

A general solution of the wave functional equation (5.5) is a linear superposition
of the particular solutions (5.6):
$$
  \psi = \sum_\pi C_\pi \psi_\pi
\eqno{(5.8)}
$$
where the sum runs over various families $\{V_4\}_1,~\{V_4\}_2$,..., defining
${\pi^a}_1$,  ${\pi^a}_1$,...\,.The
momenta are arbitrary, not restricted by a fixed ${\mathscr L} [\eta(\xi)]$.

The solution (5.8) represents a quantum state with indefinite momentum
$\pi^a [\eta (\xi)]$
and indefinite density ${\mathscr L} [\eta(\xi)]$; it is the projection of
a state $|a \rangle$  into the state $|\eta(\xi) \rangle$
with definite $\eta^a (\xi)$:
$$
  \psi[\eta(\xi),s(\xi)] \equiv \langle \eta(\xi)|a[s(\xi)] \rangle \equiv
  \psi(\Sigma,s).
\eqno{(5.9)}
$$
Here we are dealing with what I shall name the generalised Schr\"odinger
representation
in which a state depends on the proper time function $s(\xi)$.
 However, because ${\hat H} = 0$,
a state is constant at all values of $s(\xi)$. We have called 
$\psi[\eta(\xi),s(\xi)]$ the wave
functional; it is a generalisation of the concept of a wavefunction.

The wave functional has an analogous meaning as a wavefunction: if we measure
the 3-surface $\Sigma$ then the probability density that we obtain as
a result of measurement
the values $\eta^a (\xi)$ is given by:
\bear
  &&\psi^\dg [\eta(\xi)] \psi[\eta(\xi)] \nonumber \\
  &&~~~~~~~~~ = \mbox{probability density
  of finding surface $\Sigma$ with $\eta^a (\xi)$ within
 the `volume'} \nonumber\\
  &&~~~~~~~~~~~~~\mbox{element ${\mathscr D} \eta(\xi)$} \nonumber
\ear
where ${\mathscr D} \eta(\xi) = \prod_\xi \dd^N \eta(\xi)$ is the volume
element of the $\Sigma$ superspace. Normalisation of $\psi[\eta(\xi)]$
is such that
$$
  \int \psi^\dg \psi {\mathscr D} \eta = 1
\eqno{(5.10)}
$$
where the integration runs over a certain chosen volume of $\Sigma$
superspace. An explicit
meaning of the integral (5.10) is given in appendix 2 . The expression
$\psi^\dg \psi$ is the expectation value
$$
  \langle {\hat Q} \rangle \equiv \int \psi^\dg [\eta'(\xi)] {\hat Q}
  \psi[\eta'(\xi)] {\mathscr D} \eta'(\xi) 
\eqno{(5.11)}
$$
of the---let it be called---localisation operator
$$
  {\hat Q} = \delta [\eta(\xi) - \eta'(\xi)] = \prod_{\xi, a}
  \delta(\eta^a(\xi)-\eta'^a(\xi))
\eqno{(5.12)}
$$
where $\delta [~~]$ stands here for a generalised (or functional) $\delta$ 
function defined by
$$
  \int F[y'(x)] \delta [y(x)-y'(x)] {\mathscr D} y'(x) = F[y(x)]
\eqno{(5.13)}
$$
where $F[y(x)]$ is an arbitrary functional whilst $y(x)$ and 
$y'(x)$ arbitrary functions.

In general, an observable ${\hat A}$ is a functional operator and its expectation value is
given by (5.11) in which ${\hat Q}$ is replaced by ${\hat A}$. These concepts
are most easily visualised
if we consider the 3-surface $\Sigma$ as a point in the `superspace',
as already mentioned.

\section{ More about the physical interpretation of the formalism}

In the previous section we wrote down an expression for the wave functional
$\psi[\eta(\xi)] \equiv \langle \eta(\xi)|a \rangle$,
where $|a \rangle$ is a state. In general (Wheeler 1962) it is a superposition of wave
functionals with definite momenta $\pi^a (\xi)$. Now let us illustrate what we mean by a
state with definite momentum. By this we mean a family (figure 2) of spacetime surfaces
$V_4$ (or $\eta^a (x)$) each $V_4$ being a solution of the classical field equations (3.14) for a
given matter density $\omega(\eta)$. On each surface $V_4$ one can choose a three-dimensional
surface $\Sigma$ (with the parametric equation $\eta^a = \eta^a (\xi)$),
and the normal $n^\mu$ to $\Sigma$. Let us
choose a particular $V_4$ and $\Sigma$. Then we can calculate the quantity
$\pi^a(\xi)={\pi_\mu}^a n^\mu$ $= \omega \sqrt{-g} \,\p_\mu \eta^a n^\mu (\eta(\xi))$
that is the momentum density of $\Sigma$ (strictly, the momentum
conjugate to $\eta^a (\xi)$). Since it functionally depends on the chosen
$\eta^a(\xi)$, we can write $\pi^a(\xi) = \pi^a(\xi)[\eta(\xi)]$.
\setlength{\unitlength}{.8mm}
\begin{figure}[h]
\hs{3mm}
\begin{picture}(120,80)(5,-130)
\begin{turn}{-90}
\put(0,0){\includegraphics[scale=0.60]{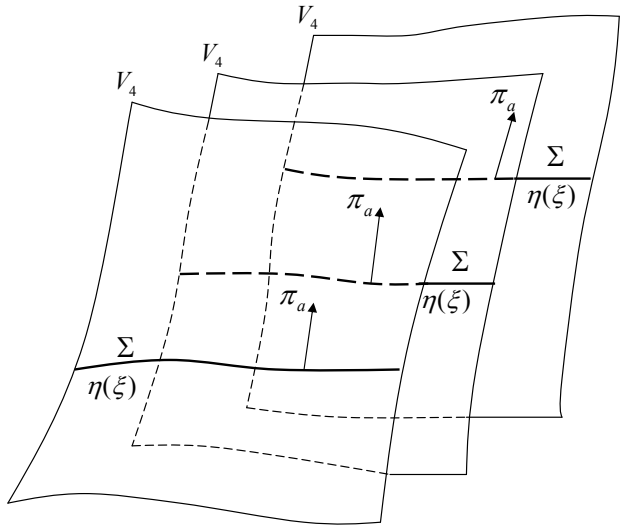}}
\end{turn}
\end{picture}

\caption{\footnotesize The representation of a state with definite momentum field
$\pi^a [\eta(\xi)]$. The figure
shows a family of the solutions $V_4$ of the field equations (3.14).
On each $V_4$ one chooses a suitable 3-surface $\Sigma$.}
\end{figure} 
Here $\pi^a (\xi)$ and $\eta^a (\xi)$ are vectors with the discrete index $a$ and
the continuous index $\xi$. If we now vary $\eta^a(\xi)$
over all spacetime surfaces $V_4$ belonging
to a given family, we can consider $\pi^a [\eta(\xi)]$
as a (generalised) $N$-vector (functional)
field ($N$ dimensions of the higher space) defined over such a set of
$\eta^a (\xi)$ (see figure 2)\footnote{
An analogous definition of momentum field for the case of a family of worldlines in a fixed
background metric is given by Pav\v si\v c (1982).}.
For a state with a definite momentum it is the field $\pi^a (\xi)[\eta^a (\xi)]$
which is known
to an observer, whilst the surface $\Sigma$ is not known. It could be any $\Sigma$
of the family $\lbrace V_4 \rbrace$
defining $\pi^a$. Just the opposite is the situation in which the observer gets as a result
of measurement the definite surface $\Sigma$ (and the matter distribution
$\rho (x) \equiv \omega (\eta(x))$ on 
this 3-surface) whilst he has no information about the family of surfaces $V_4$, i.e. about
`alternative histories'. There are also situations between these two extremes: they are
represented by a certain superposition of the states with definite momentum field
$\pi^a [\eta(\xi)]$ (figure 2).

The examples described above are in fact manifestations of (generalised) Heisenberg
uncertainty relations. They result from the following commutation relations:
$$
  [{\hat \eta}^a, {\hat \pi}_b] = i {\delta^a}_b
\eqno{(6.1)}
$$
$$
  [{\hat s}, {\hat {\mathscr L}}] = i
\eqno{(6.2)}
$$
One can easily prove these relations in the `coordinate' representation in which the
operators ${\hat \eta}^a$ are c-number fields $\eta^a (\xi)$ and $s(\xi)$,
respectively, whilst the operators
${\hat \pi}_b$ and ${\hat {\mathscr L}}$ are the functional derivatives
$$
  {\hat \pi}_a = -i \frac{\delta}{\delta \eta^a (\xi)}~~~~~~~~~{\hat {\mathscr L}} 
  = -i \frac{\delta_{\rm T}}{\delta s(\xi)}
$$
Actually
$$
  \left [ \eta^a, - i \frac{\delta}{\delta \eta^b} \right ] \!\psi =
  \eta^a \left ( - i \frac{\delta \psi}{\delta \eta^b} \right ) + 
  i \frac{\delta (\eta^a \psi)}{\delta \eta^b} = i {\delta^a}_b \,\psi
$$
and similarly for the relation (6.2). Once verified in one representation they must be
true in any representation.

The physical meaning of the commutation relations (6.1)
and (6.2) is the following.

    (a) {\it Definite}  $\pi_a [\eta(\xi)]$; then also
${\mathscr L}^2 = \pi^a[\eta(\xi)]\pi_a[\eta(\xi)]$ and the family of
spacetime surfaces $\eta^a (x)$ are {\it definite}. On the other hand,
the individual 3-surface $\eta^a (\xi)$ 
and the proper time field $s(\xi)$ are {\it indefinite}.

    (b) {\it Definite} individual $\eta^a (\xi)$; then
$\pi^a [\eta(\xi)]$ is {\it indefinite}---an observer has no
information about other spacetime surfaces. On the other hand,
the matter distribution
field ${\mathscr L}[\eta(\xi)]$ can be: (i) either {\it definite};
then the proper time field $s(\xi)$ is {\it indefinite},
or (ii) {\it indefinite}; then $s(\xi)$  can be {\it definite},
as suggested by (6.2).

    A 3-surface $\eta^a (\xi)$ and the proper time field $s(\xi)$
define a set of various spacetime
surfaces $V_4$, passing through that particular 3-surface $\Sigma$,
all $V_4$ having the same metric
$g_{\mu \nu}(x) $ (apart from a general coordinate transformation).

The commutation relations (6.2) can also be interpreted in the following way: if
an observer obtains as a result of measurement a definite metric, then he has no
knowledge about the matter distribution field ${\mathscr L}[\eta(\xi)]$,
i.e. he has no information
about the matter distribution on an alternative three-dimensional
surface $\Sigma$; therefore
he does not know alternative worlds. To state it differently:
if an observer measures
the proper time $s(\xi)$ in each point $\xi$ of his 3-space, then he cannot
also measure with
an arbitrary precision the matter distribution field 
${\mathscr L} (\xi)={\mathscr L}(\xi)[\eta(\xi)] = \rho[\eta(\xi)] \sqrt{-g}$.

    In a normal, awake, state an observer is continuously measuring (at least with his
sense organs) the proper time; he has no idea of 'other worlds'. Besides this situation,
our theory also predicts a situation in which an observer does not measure the proper
time, and as a compensation he can then have some knowledge about the distribution
of matter through the higher space, i.e. he can experience in some way the existence
of alternative worlds. Has this last prediction any relation to reality? My conjecture
is that it has. Remember that an observer is not always in an awake state of
consciousness. He can be under the influence of hypnosis, drugs, etc, or he can simply sleep
and dream. When in such a state of consciousness, he is no longer precisely `measuring'
the proper time; he is not aware of the usual three-dimensional space or world, but
nevertheless he is aware of something: he experiences various halucinations, dreams,
etc. Is this related to some certain extent with the existence of higher dimensional
space and alternative worlds? If so, then the fact that these experiences are often not
strictly logical or causal can be a consequence of the fact that in such a state the
observer is not measuring (at least not precisely enough) his proper time.

\section{Transition to the one-particle theory}

So far we have been concerned with the classical and quantum motion of a 3-surface
$\Sigma$. Classically, a motion of $\Sigma$ gives an observer the impression that
three-dimensional
objects---the sections of higher dimensional objects with $\Sigma$---are moving in 3-space.
Classical motion of $\Sigma$ is only a limiting case of a more general, quantum motion, for
which it is characteristic that one cannot determine at once both $\Sigma$ itself and the
momentum $\pi^a = \omega \, \p_\mu \eta^a n^\mu \srg$.
Since the motion of $\Sigma$ is due to quantum uncertainty
the motion of material 3-objects is also quantum mechanically uncertain.

Let us derive the wavefunction of a 3-space material particle from the wave
functional $\psi(\Sigma)$. The latter is a superposition of the wave functionals with definite
momentum $\pi^a$ . Let the matter distribution $\omega (\eta)$ in higher space be
$$
  \omega (\eta) = m \delta^N (\eta - \eta_0 (\alpha))
\eqno{(7.1)}
$$
i.e. the matter in higher space is distributed on a given four-dimensional surface $V_4^*$,
$\eta^a = \eta_0^a (\alpha)$ ($\alpha$ are the coordinates on $V_4^*$), and is zero elsewhere.
The state with a
definite momentum field $\pi^a$ can be represented by a family of spacetime surfaces $V_4$
(each being a solution of the field equations (3.14) for a given $\omega (\eta)$).
Each $V_4$ intersects
with $V_4^*$ in a one-dimensional continuum---a worldline ${\tl C}$. If we project all those
worldlines on a certain spacetime slice $V_4^0$, we obtain a family of possible worldlines
$P$ in a given spacetime (in our case $V_4^0$).

Without loss of generality we can assume that $n^\mu = u^\mu$ (choice of $\Sigma$ on 
$V_4^0$) and write
$$
  \pi_a \delta \eta^a (\xi) = \pi_a \,\dd \eta^a = \srg \,\omega U_a \p_\mu \eta^a \dd x^\mu 
  = \srg \, \omega u_\mu \, \dd x^\mu
\eqno{(7.2)}
$$
where $\dd \eta^a$ is the projection of $\delta \eta^a (\xi)$ into $V_4^0$.
Here $\delta \eta^a (\xi)$ is taken between those
points on $V_4$ and $V_4+ \delta V_4$ which lie on the corresponding worldlines (figure 3).
From (7.2) one obtains
$$
  \iint \pi_a \delta \eta^a (\xi) \dd \Sigma = \int p_\mu \dd x^\mu ~~~~~~~~~~~~~~
  p_\mu = \int \srg \, \omega \, u_\mu \, \dd \Sigma .
\eqno{(7.3)}
$$

\setlength{\unitlength}{.8mm}

\begin{figure}[h!]
\hs{3mm}
\begin{picture}(120,80)(20,-92)
\begin{turn}{-90}
\put(0,0){\includegraphics[scale=0.60]{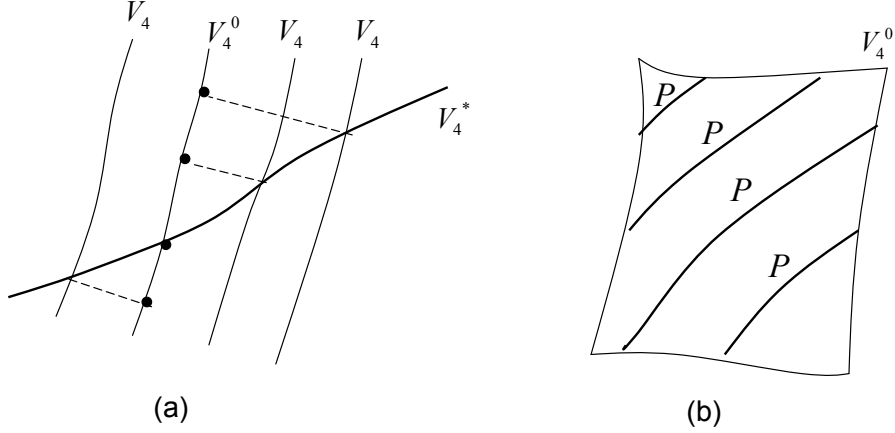}}
\end{turn}
\end{picture}

\caption{\footnotesize (a) A family of spacetime slices $V_4$ intersected by a surface $V_4^*$ 
of non-vanishing
matter distribution $\omega (\eta)$. (b) The intersections $V_4 \cap V_4^*$ are projected
on a chosen spacetime
surface $V_4^0$,   thus giving a family of possible worldlines $P$.}
\end{figure} 
\nnn Therefore the wave functional (5.6) actually becomes in this special case the
wavefunction
$$
  \psi_p (x) = \psi_0 \,{\rm exp} \left ( i \int p_\mu \dd x^\mu \right ) .
\eqno{(7.4)}
$$
A general wavefunction is a superposition of $\psi_p$ and obeys the wave equation\footnote{
{\bf Note added in 2014}: In a curved space $V_4^0$, the spin connection should also
occur in eqs.\,(7.5) and (7.6). This can be brought into the game, if
$\psi$ is a Clifford algebra valued (geometric) spinor, expanded according to
$\psi = \psi^\alpha \xi_\alpha$, $\alpha =1,2,3,4$, where $\xi_\alpha$ are basis
spinors. The derivative $\p_\mu$ can be understood in the generalised sense, such that
if acting on a scalar component it gives the partial derivative $\p_\mu \psi^\alpha$,
and if acting on a basis spinor it gives 
$\p_\mu \xi_\alpha = \Gamma_{\mu ~ \alpha}^{~\beta} \xi_\beta$, where 
$\Gamma_{\mu ~ \alpha}^{~\beta}$ is the spin connection. Then $\p_\mu \psi =
\p_\mu (\psi^\alpha \xi_\alpha)$ $= \p_\mu \psi^\alpha \xi_\alpha + \psi^\alpha
\p_\mu \xi_\alpha$ $=(\p_\mu \psi^\alpha + \Gamma_{\mu ~ \beta}^{~\alpha} \psi^\beta) \xi_\alpha
\equiv (\DD_\mu \psi^\alpha) \xi_\alpha$ (see M. Pav\v si\v c 2006 {\it Int. J. Mod. Phys.} A
{\bf 21} 5905--56). }
$$
  {\hat L} \psi = \gam^\mu {\hat p}_\mu   \psi~~~~~~~~ (\,{\hat p}_\mu = - i \p_\mu ,
  ~~~~~~~~ {\hat L} = - i \dd/\dd s \,).
\eqno{(7.5)}
$$
For a state with definite mass $m$ it is ${\hat L} \psi = m \psi$,
and equation (7.5) becomes
$$
  (\gam^\mu {\hat p}_\mu  - m) \psi = 0
\eqno{(7.6})
$$
which is the well known Dirac equation in a fixed background metric.

   These results essentially mean that starting from a quantum theory of a spacetime
surface embedded in a higher dimensional space with a given matter distribution $\omega (\eta)$,
one obtains in the special case (given by (7.1)) the one-particle quantum theory. For
an arbitrary (in general complicated)  $\omega (\eta)$) one would obtain many-particle quantum
theory. Both quantum gravity (at least the model theory presented here) and quantum
theory of matter are thus intimately related\footnote{
Here we neglect other forces, like the electromagnetic, strong, etc. The simultaneity surface $\Sigma$
is three-dimensional only as a working hypothesis, in order to reproduce the usual, four-dimensional
Einstein gravity.
If instead, we add some more dimensions to $\Sigma$, we could reproduce the higher dimensional gravity which,
according to Kaluza (1921) and Klein (1926, 1928) and the modern elaborations (Luciani 1978), includes
other forces.}.

\section{ Conclusion}

We have developed what appears to be a consistent formalism of a model theory for
quantum gravity, based on the quantisation of a spacetime surface embedded in a
higher dimensional space $V_N$; namely, we consider the space ${\mathscr S}=\lbrace \Sigma
\rbrace$ of the simultaneity
3-surfaces $\Sigma \subset V_N$, and suitably fix its measure. Each $\Sigma$ can be considered as a point
in the superspace ${\mathscr S}$.  This superspace then represents the basis for the Hilbert space
of functions $f(\Sigma)$. A quantum state is represented by a vector in this Hilbert space.
In the classical approximation an observer experiences that $\Sigma$ proceeds forward in his
proper time and describes a four-dimensional spacetime continuum $V_4$. In our theory
various spacetime surfaces coexist in $V_N$ and they represent different classical histories
of events. Because of quantum effects there is a certain interference between different
histories, as described in \S\,6. Roughly speaking, one has the uncertainty principle
between metric (or proper time) and matter distribution in the higher space (or
alternative histories). So an observer may face a state in which he has complete
knowledge about the metric, proper time and matter distribution in his simultaneity
3-surface, but absolutely no knowledge about alternative histories or spacetimes and
the matter distribution on them. This is the state we are all familiar with. On the other
hand, the uncertainty principle of \S\,6 implies that an observer can experience a state
in which he has no knowledge (or not precise enough knowledge) of the proper time
and matter distribution on a certain simultaneity 3-surface, but instead he has some
knowledge about alternative histories situated on different spacetime sheets.
In particular, in the usual quantum mechanical experiments, an observer, who is not
continuously measuring a particle's position, obtains certain knowledge about the particle's
alternative positions (or histories); namely, the very existence of an interference pattern
in the double slit experiment is then a manifestation of alternative possible positions
of the particle, formally represented by its wavefunction. These alternative positions
lie on different spacetime slices and belong to the corresponding different (classical)
histories; the usual quantum mechanical situations (e.g. one-particle motion) are
limiting cases of our generalised quantum theory of a spacetime sheet, such that the
metric tensor is not fluctuating but remains the same on all spacetime slices (see \S\,7) .

In our theory we have a very interesting link between determinism and indeterminism.
Deterministic or given is the matter distribution in higher space; this higher
dimensional world is fixed and timeless---this is the physical reality. On the other
hand, the world as perceived by an observer is indeterministic: the path of an observer's
three-dimensional `now' is unpredictable and obeys the quantum laws as described in
\S\S\,5--7. A certain path is only one of many possible paths through the higher space.
The world as perceived by an observer on a certain spacetime sheet is not the only
possible world. There are other worlds (and other observers)---all belonging to the
higher dimensional world---which are not directly perceived or measured by an observer.
Their existence manifests themselves to an observer through quantum phenomena.

The well known quantum phenomena are just a subset of phenomena belonging
to the proposed theory of spacetime $V_4$. We have not fully explored the experimental
consequences and predictions of the proposed theory but only set its conceptual and
formal foundations, trying to demonstrate its self-consistency and usefulness in unifying
various branches of physics. What we present here should be regarded for the moment
only as a model theory for quantum gravity. We hope that physically relevant solutions
could be found non-perturbatively, thus avoiding the problem of renormalisability.
We suggest future work in two directions: (i) to evaluate the experimental consequences
of the theory regarding the gravity itself and eventually find a more realistic Lagrangian
and (ii) to generalise the theory to more than the four-dimensional spacetime surface,
thus bringing into play---via the Kaluza-Klein mechanism---other interactions besides
gravity.

\vs{5mm}

{\bf Appendix 1. Definition of the phase functional}

\vs{3mm}

\nnn In \S\,4 we have defined the generator of infinitesimal variations of $V_4$
$$
  \delta S = \delta S_1 + \delta S_2 = \int (\pi_a \delta \eta^a (x) +
  {\mathscr L} \delta s ) \, \dd \Sigma = \int \pi_a {\bar \delta} \eta^a (x) \, \dd \Sigma
\eqno{(A1.1)}
$$
where ${\bar \delta} \eta^a = \eta'^a (x')-\eta^a (x)$, $\delta \eta^a (x) =
\eta'^a (x) - \eta^a(x)$ and ${\mathscr H} = 0$. We shall define
the phase functional by the integral
$$
  S(V_4, s) = S_1 + S_2 = \iint_{C,\Sigma} \pi_a {\bar \delta} \eta^a (x) \, \dd \Sigma .
\eqno{(A1.2)}
$$
Here $V_4$ is a spacetime surface also denoted by $V_4 (\alpha) =
\eta^a (x,\alpha)$, where $\alpha$ is a set of
parameters (two parameters) which, when fixed, determine a particular $V_4$; the latter
is a solution of the second-order equation (3.14). Now we let $\alpha$ vary. Thus we obtain
a family $F = \{ V_4(\alpha) \}$
(figure 1$b$ ) and the set
$$
   {\mathscr S}_F = \lbrace \eta^a (\xi) \rbrace_F = \{ \eta^a (\xi): 
   \eta^a (\xi) \subset V_4 (\alpha), ~V_4 (\alpha) \in F \}.
\eqno{(A1.3)}
$$
Next we choose a path $C=\{\eta(\xi) \}_C$ (used in (A1.2)) which is a subset of
${\mathscr S}_F$ (figure l$b$), consisting of $\eta^a (\xi)$ such that each
$\eta^a (\xi) \in C$ belongs to different $V_4 (\alpha) \in F$. The
integral (A1.2) is the limit
$$
  S = \lim_{h \to o} \left ( \int_\Sigma \pi_a {\bar \delta} \eta^a (x)
  \dd \Sigma|_{V_4 (\alpha)} 
  + \int_\Sigma \pi_a {\bar \delta} \eta^a (x)
  \dd \Sigma|_{V_4 (\alpha +h)} + ... \right )
\eqno{(A1.4)}
$$
where ${\bar \delta} \eta^a (x)$ is chosen along the path $C$ crossing various
$V_4 (\alpha)$ (figure l$b$). The
definition of ${\bar \delta} \eta^a (x)$ is already given in (Al.l) and $\S$ 4,
but it is instructive to clarify it as follows.

Since $\delta \eta^a = \eta'^a (x) - \eta^a (x)$ is a variation of
$\eta^a (x)$ at fixed $x^\mu$ it is also $\delta \eta^a
= [\eta'^a (\xi) - \eta^a (\xi)]_\bot \equiv \delta \eta_\bot^a (\xi)$,
i.e. a variation of the 3-surface $\eta^a (\xi)$, `normal' to $V_4$.
Now, using $u_\mu = U_a \p_\mu \eta^a$, $\pi_a = \srg \, \omega U_a$ and
$\delta \eta_\Vert^a \equiv \p_\mu \eta^a \delta x^\mu$, the term
${\mathscr L} \delta s$  in (Al.l) can be written as
$$
  {\mathscr L} \delta s = \srg \, \omega u_\mu \delta x^\mu
   = \pi_a \delta \eta_\Vert^a .
\eqno{(A1.5)}
$$
Hence the total variation ${\bar \delta} \eta^a (x)$ is equal to the variation
$\delta \eta^a (\xi)$ and is the sum of
the normal variation $\delta \eta_\bot^a (\xi)$ and the tangent variation
$\delta \eta_\Vert^a (\xi)$ :
$$
  {\bar \delta} \eta^a (x) = \delta \eta^a (\xi) =
  \delta \eta_\bot^a (\xi) + \delta \eta_\Vert^a (\xi) .
\eqno{(A1.6)}
$$
The phase functional (A1.2) is the sum of the term $S_1$ due to a normal and the term
$S_2$ due to a tangent variation.

The phase functional can also be understood as the limiting case---for continuum
$\xi_i$---of the equation
$$
  S(\eta_A (\xi_i),\eta_B (\xi_i))
   = \sum_k \int_A^B \pi_a \, \dd \eta'^a (\xi_k) \Delta \Sigma^k
\eqno{(A1.7)}
$$
where $A$ denotes an initial 3-surface and $B$ a final one. Before the limiting procedure
is performed, $\eta^a (\xi)$ is approximated by a discrete set of points
$\eta^a (\xi_i) \equiv (\eta^a (\xi_1), \eta^a (\xi_2),...)$
which can be considered as a vector in a finite (or infinite but
countable) dimensional space ${\mathscr S}_{n F}$.

From (A1.7) it is obvious that $\pi_a (\eta (\xi_i))
= \p S/\p \eta^a (\xi_i) (\Delta \Sigma_i)^{-1}$.
Therefore $S(\eta_A (\xi_i),\eta_B (\xi_i))$ is
independent of the path between the points $A$ and $B$ in ${\mathscr S}_{n F}$,
and---for a fixed $\eta_A$---it is a unique function $S(\eta (\xi_i))$
(we omit the subscript $B$).

In the limit of continuum $\xi_i$ we obtain the unique phase functional $S[\eta(\xi)]$ . The
momentum is the functional derivative $\pi_a [\eta (\xi)]$ $= \delta S/\delta \eta^a (\xi)$.
For different families
$F = \{V_4(\alpha)\}$ we obtain different $S_F [\eta (\xi)]$, different spaces
${\mathscr S}_F = \{\eta (\xi)\}_F$ and different momenta $\pi_a [\eta (\xi)]_F$.

\vs{5mm}

{\bf Appendix 2. Fixation of the measure in $\Sigma$ superspace}

\vs{3mm}

\nnn We are interested in the set of 3-surfaces $\Sigma \equiv \eta^a (\xi)$
which can be expanded over a certain orthonormal set of functions
${P_n}^a (\xi)$:
$$
  \eta^a (\xi) = \sum_{n=0}^\infty {\alpha_n}^a {P_n}^a (\xi)
  ~~~~~(\mbox{no sum over}~a).
\eqno{(A2.1)}
$$
Let ${P_0}^a (\xi) =1$. We do not fix the `end points' so that (A2.1) also includes the translated
3-surfaces (determined by the choice of ${\alpha_0}^a$) . ${P_n}^a (\xi)$ for $n = 1,2,...,\infty$ represent a
basis in a function space---which we call $\Sigma$ superspace ${\mathscr S}_P$.
A choice of some other non-equivalent basis, say $Q_n^a$,
would span a different superspace ${\mathscr S}_Q$.

A variation of a 3-surface belonging to ${\mathscr S}_P$ is
$$
  \delta \eta^a (\xi) = \sum_{n=0}^\infty 
  \frac{\p \eta^a}{\p {\alpha_n}^a} \dd {\alpha_n}^a
  = \sum_{n=0}^\infty {P_n}^a \, \dd {\alpha_n}^a  ~~~~~(\mbox{no sum over}~a).
\eqno{(A2.2)}
$$
The measure ${\mathscr D} \eta (\xi)$ in the space of $\eta^a (\xi) \in {\mathscr S}_P$
is defined as
$$
  {\mathscr D} \eta (\xi) = \prod_{\xi,a} \dd \eta^a (\xi_i) =
  \frac{\p(\eta^a)}{\p(\alpha)} \prod_{n=0}^\infty \prod_{\alpha=1}^N \dd {\alpha_n}^a
\eqno{(A2.3)}
$$
where $\p(\eta^a)/\p(\alpha) = {\rm det} \, {P_n}^a (\xi_i)$ is the Jacobian of the
transformation (A2.1) from the variables $\eta^a (\xi_i)$ variables ${\alpha_n}^a$.

For the expectation value of a generic operator ${\hat A}$ we have
$$
  \langle {\hat A} \rangle 
  = \int \psi^\dg [\eta [\xi)] {\hat A} \psi [\eta (\xi)] {\mathscr D} \eta \hs{6.5cm}
$$
$$
  = \int \psi^\dg (\alpha_1,\alpha_2,...) {\hat A} \psi (\alpha_1,\alpha_2,...)
  \frac{\p(\eta^a)}{\p(\alpha)} \prod_{n=0}^\infty \prod_{\alpha=1}^N \dd {\alpha_n}^a .
\eqno{(A2.4)}
$$

As an example let us calculate the probability $w$ of finding as a result of measurement
a certain $\eta^a (\xi)$ represented by (A2.1) within a given volume of the superspace
${\mathscr S}_P$. Let the wave functional be a superposition of $\psi_\pi$
(equation (5.6)):
$$
  \psi[\eta(\xi)] = \sum_k c_k N {\rm exp} \left ( i \iint \pi_a^{\,(k)} \delta \eta^a
   \, \dd \Sigma \right ) \hs{2.5cm}$$
$$
  = \sum_k c_k N {\rm exp} \left ( i \sum_n {\mathscr P}_{n a}^{(k)} ( {\alpha_n}^a -
  {\alpha_n}^a (0)) \right )
\eqno{(A2.5)}
$$
where ${\mathscr P}_{n a}^{(k)} \equiv \int \pi_a P_{na}(\xi) \, \dd \Sigma$
(no sum over $a$). Then using (A2.3), identifying
${\bar N}^\dg {\bar N} \equiv N^\dg N \p (\eta)/\p (\alpha)$
and setting ${\alpha_n}^a (0) = 0$ it is
$$
  w = \int \psi^\dg \psi {\mathscr D} \eta = \int \sum_{k,k'} c_k^* c_{k'}
  \prod_n {\bar N}^\dg {\bar N} \, {\rm exp} \left (i {\alpha_n}^a
  ({\mathscr P}_{na}^{(k)} -{\mathscr P}_{na}^{(k')}) \right ) \dd^N \alpha \hs{1cm}$$
$$
  = \mu \sum_{k,k'} c_k^* c_{k'} \prod_n \delta^N
   ({\mathscr P}_{na}^{(k)} -{\mathscr P}_{na}^{(k')})) 
  = \sum_{k k'} c_{k}^*c_{k'} \delta_{k k'} = \sum_k c_k^* c_k .
\eqno{(A2.6)}
$$
where $\mu$ is a suitable constant which absorbs various normalisation factors.

The question is what determines the superspace ${\mathscr S}_P$. The latter is a subspace of the
space of all possible configurations $\eta^a (\xi)$. We argue that the choice
of the basis $P$ and hence of ${\mathscr S}_P$ is implicit in the construction
of the measuring apparatus. By the
latter we understand a structure in $V_N$ consisting of the chain:
sensor-brain, where sensor is a chain: artificial sensor-instrument-sense organ.
This is in a sense analogous
to the usual quantum mechanics where a set of eigenvalues is determined by a kind
of measuring device (not necessarily identical with the measuring apparatus defined
above). In other words, as a measuring device determines which set of eigenvalues
we shall measure, so the measuring apparatus, by fixing ${P_n}^a (\xi)$,
determines which set $\{ \eta (\xi) \}_P \equiv {\mathscr S}_P$ is disposable.

\vs{4mm}

{\bf References}

\vs{3mm}

{\footnotesize \baselineskip .45cm
{\obeylines
Aghassi J J, Roman P and Santilli R M 1970 {\it Phys. Rev.} D {\bf 10} 2753--65
Anderson J L 1964 {\it Gravitation and Relativity}  ed Hong-Yee Chin and W F Hoffman (New York:
~~~~~~     Benjamin) p 279
Ashtekar A and Geroch R 1974 {\it Rep. Prog. Phys.} {\bf 37} 1211--56
Barut A O and Mullen G H 1962 {\it Ann. Phys., NY} {\bf 20} 203--18
------ 1964 {\it Electrodynamics and Classical Theory of Fields and Particles} (New York: Macmillan)
~~~~~~~~p 122
Blokintsev D 1973 {\it Space and Time in the Microworld} (Dordrecht: D Riedel) p 147
Bohr H and Nielsen H B 1983 {\it Nucl. Phys.} B {\bf 212} 547--55
Brill D R and Gowdy R H 1970 {\it Rep. Prog. Phys.} {\bf 33} 413-88
Clarke C J S 1970 {\it Proc. R. Soc.} A {\bf 314} 417--28
De Beauregard 0 C 1979 in {\it The Iceland Papers, Frontiers of hysics Conference, Reykjavik 1977}
~~~~~~~~(London: Essential Research Associates)
DeWitt B S 1967a {\it Phys. Rev.} {\bf 160} 1113--4
------ 1967b in {\it Battelle Rencontres 1 (Lectures in Mathematics and Physics)} (New York: 
~~~~~~~~Benjamin) p 319
Eisenhart L P E 1926 {\it Riemannian Geometry} (Princeton: Princeton University Press)
Everett H 1957 {\it Rev. Mod. Phys.} {\bf 29} 454--62
Fradkin E S and Tsetlin A A 1982 {\it Ann. Phys., NY} {\bf 143} 413--47
Fronsdal C 1959 {\it Nuovo Cimento} {\bf 13} 988--1006
Gervais J L and Neveu A 1982 {\it Nucl. Phys.} B {\bf 209} 125--45
Henneaux M 1983 {\it Phys. Lett.} B {\bf 120} 179-82
Horwitz L P and Arshanski R 1982 {\it J. Phys. A: Math. Gen.} {\bf15} L659--62
Horwitz L P and Piron C 1973 {\it Helv. Phys. Acta} {\bf 46} 316--26
Kaluza T 1921 {\it S. B. Preus. Akad. Wiss., Math. Naturw.} {\bf K1} 966
Klein 0 1926 {\it Z. Phys.} {\bf 37} 895
------ 1928 {\it Z. Phys.} {\bf 46} 188
Kato M and Ogawa K 1983 {\it Nucl. Phys.} B {\bf 212} 443--60
Landau L D and Lifshitz E M 1967 {\it Teorija polja} (Moscow: Nauka) p 430
Luciani J F 1978 {\it Nucl. Phys.} B {\bf 135} 111--30
Menski M B 1976 {\it Commun. Math. Phys.} {\bf 47} 97--108
Pav\v si\v c M 1981a {\it J. Phys. A : Math. Gen.} {\bf 14} 3217--28
------ 1981b {\it Lett. Nuovo Cimento} {\bf 30} 111--5
------ 1982 {\it Phys. Lett.} A {\bf 9} 175--7
------ 1984 {\it Nuovo Cimento} A {\bf 82} 443--55, A 83 305
------ 1985 {\it Phys. Lett.} A {\bf 107} 66--72
Penrose R 1965 {\it Rev. Mod. Phys.} {\bf 37} 215
Polyakov A M 1981a {\it Phys. Lett.} B {\bf 103} 207--10
------ 1981b {\it Phys. Lett.} B {\bf 103} 211-3
Rund H 1971 {\it Invariant Theory o Variational Problems on Subspaces of a Riemannian Manifold}
~~~~~~~~(G\"ottingen: Vandenhoock Rupret)
Stueckelberg E C G 1941a {\it Helv. Phys. Acta} {\bf 14} 316
------ 1941b {\it Helv. Phys. Acta} {\bf 15} 23--37
------ 1941c {\it Helv. Phys. Acta} {\bf 15} 51--5
Tataru-Mihai P 1982 {\it Nuovo Cimento} A {\bf 72} 80--5
Wheeler J A 1957 {\it Rev. Mod. Phys.} {\bf 29} 463--6
------ 1967 in {\it Battelle Rencontre I (Lectures in Mathematics and Physics)} (New York:
~~~~~~~~Benjamin) pp 242--307
------ 1973 {\it The Physicist's Conception o Nature} ed J Mehra (Dordrecht: D Reidel) p 202
Wigner E P 1967 {\it Symmetries and Reflections} (Bloomington: Indiana University Press)
~~~~~~~~pp 171--207
}}

\end{document}